\documentclass[aip,prb,amsmath,amssymb,reprint]{revtex4-1}
\usepackage{graphicx}
\usepackage[version=3]{mhchem} 
\usepackage{subfig}
\usepackage{dcolumn}
\usepackage{bm}

\begin{document}

\title{Relativistic quasiparticle self-consistent electronic structure of hybrid halide perovskite photovoltaic absorbers}

\author{Federico Brivio}
\affiliation{Centre for Sustainable Chemical Technologies and Department of Chemistry, University of Bath, Claverton Down, Bath BA2 7AY, UK}

\author{Keith T. Butler}
\affiliation{Centre for Sustainable Chemical Technologies and Department of Chemistry, University of Bath, Claverton Down, Bath BA2 7AY, UK}

\author{Aron Walsh}
\email[Electronic mail:]{a.walsh@bath.ac.uk}
\affiliation{Centre for Sustainable Chemical Technologies and Department of Chemistry, University of Bath, Claverton Down, Bath BA2 7AY, UK}

\author{Mark van Schilfgaarde}
\email[Electronic mail:]{mark.van$_$schilfgaarde@kcl.ac.uk}
\affiliation{Department of Physics, Kings College London, London WC2R 2LS, UK}

\date{\today}

\begin{abstract}
  Solar cells based on a light absorbing layer of the organometal
  halide perovskite CH$_3$NH$_3$PbI$_3$ have recently surpassed 15
  \% conversion efficiency, though how these materials work remains
  largely unknown.
  We analyse the electronic structure
  and optical properties within the quasiparticle self-consistent
  \emph{GW} approximation.  While this compound bears some
  similarity to conventional \emph{sp} semiconductors, it also
  displays unique features.  Quasiparticle self-consistency is
  essential for an accurate description of the band structure:
  band gaps are much larger than what is predicted by the local
  density approximation (LDA) or \emph{GW} based on the LDA.
  Valence band dispersions are modified in a very unusual manner.
  In addition, spin-orbit coupling strongly modifies the band
  structure and gives rise to unconventional dispersion
  relations and a Dresselhaus splitting at the band edges.  
  The average hole mass is small, which partially
  accounts for the long diffusion lengths observed. 
  The surface ionisation potential (workfunction) is calculated to be
  5.7 eV with respect to the vacuum level, explaining efficient carrier
  transfer to TiO$_2$ and Au electrical contacts.
\end{abstract}


\maketitle

\section{Introduction}
One of the most promising third-generation photovoltaic technologies
is based on metal-organic halide
perovskites.\cite{kojima-6050,snaith-1,gratzel-1,gratzel-2,heo-486,snaith-2,snaith-6156,snaith-7467,gratzel-6156,bisquert-4}
The materials physics of inorganic (ABX$_3$) perovskites is well
developed; however, the replacement of the inorganic cation by an
isoelectronic organic moiety provides an opportunity for tuning the
chemical bonding and optical response.
We apply a range of electronic structure techniques to calculate 
and predict the band structure of hybrid perovskites, 
demonstrating how the rich and unusual 
physics of these materials accounts for their widely reported success 
as absorber layers in solar cells.

It has been established that similar to traditional dielectric
perovskites, these hybrid analogues have a range of accessible
polymorphs with variations in the tilting and rotation of the BX$_6$
polyhedra in the lattice.\cite{baikie-5628} 
A large family of hybrid
perovskites have been reported with inorganic networks ranging 
from 1--3 dimensions.\cite{calabrese-2328,mitzi-1,mitzi-2,borriello-235214}
However, the methylammonium (MA) cation (\textit{i.e.} CH$_3$NH$_3^+$)
has been widely applied, resulting in the highest-performance
perovskite-structured solar absorbers.\cite{snaith-1,gratzel-1}
The polar MA cation can also be replaced by ammonium (NH$_4^+$) as a
smaller non-polar analogue.

A large number of density functional theory (DFT) studies have been reported that
examine the electronic properties of hybrid perovskites.\cite{borriello-235214,mosconi-13902,quarti-279,brivio-042111,yin-063903,filippetti-125203,even-2999,giorgi-4213}
The majority neglect spin-orbit coupling,\cite{borriello-235214,mosconi-13902,quarti-279,brivio-042111,yin-063903,filippetti-125203}
while a relativistic treatment based on local or semi-local exchange-correlation 
functionals results in severe band gaps underestimations.\cite{even-2999,giorgi-4213}   
Both approaches are insufficient to describe the complexity of the electronic
structure of these hybrid semiconductors, with large errors expected in predicted properties
such as carrier effective mass and dielectric function. 

An alternative approach is the $GW$ formalism, 
which can be used to correct errors in
the one-electron Kohn-Sham eigenvalues within a many-body
quasiparticle framework.
Here we employ quasiparticle self-consistent $GW$
theory\cite{mark06qsgw} (QS\emph{GW}) to study the electronic
structure of CH$_3$NH$_3$PbI$_3$ and NH$_4$PbI$_3$, including the
effect of spin-orbit coupling (SOC) $\lambda \mathbf{L}{\cdot}\mathbf{S}$, on
the both the kinetic energy and electron self-energy $\Sigma$
(see the Appendix).  As Pb and I are heavy elements, SOC is large
and has a major effect on spectral properties. SOC predominantly
modifies the kinetic energy; however, in this case relativistic
effects are large enough to induce a modest reduction in $\Sigma$
as well, in contrast to the vast majority of semiconductors,
e.g. elemental Sn.  As a consequence of large relativistic
effects, the conduction and valence bands near the band extrema
deviate strongly from parabolic behavior.  Effective masses are
no longer constant, but depend on doping, temperature, and the
property being measured.  Average effective masses are
nevertheless light, and the dielectric constants large, 
 accounting for the long diffusion
lengths that have been recently reported.\cite{snaith-6156,gratzel-6156}

In many respects these perovskites are similar to conventional
\emph{sp} semiconductors: conduction and valence bands near the
Fermi level have \emph{sp} character, and local- (and
semi-local-) density approximation (LDA) to DFT 
systematically underestimate the band gap $E_G$
because they do not include spatial non-locality in the
exchange-correlation potential.  There are other significant
points of departure: in sharp contrast to tetrahedral
semiconductors, DFT also poorly describes valence band
dispersions.  This surprising result, which we discuss further
below, indicates that the usual explanations invoked to account
for deficiencies in DFT's description of semiconductors are not
sufficient here.

We show that there is a strong feedback between dielectric
response and quasiparticle levels, as occurs for CuInSe$_{2}$
\cite{vidal-056401}.  
Thus self-consistency in \emph{GW} is essential: $E_G$
calculated from $G^\mathrm{LDA}W^\mathrm{LDA}$,
i.e. LDA as the starting Hamiltonian, picks up only a little better
than half the gap correction to the LDA.  Moreover, the
QS\emph{GW} and LDA valence bands, which the LDA describes
reasonably well in tetrahedral semiconductors, are significantly
different.  These differences underscore the limitations of
density-functional based approaches (LDA, hybrid functionals, or
$G^{\rm LDA}W^{\rm LDA}$) in describing the properties of these
materials. 
QS\emph{GW} does not depend on the LDA:
self-consistency renders it more reliable and universally
applicable than other forms of $GW$, which will be important 
for \textit{in silico} design of hybrid systems.  
Moreover, QS\emph{GW} can
determine some ground-state properties, e.g. the charge density
and electric field gradient.  Errors in QS\emph{GW} tend to be
small and highly systematic; most notably there is a tendency to
slightly overestimate semiconductor band gaps.  Limited data is
available for organic-inorganic halide perovskites, but at least
for CH$_3$NH$_3$PbI$_3$ the universal tendency found in other
materials is consistent with recent measurements.

Finally, based on the workfunction calculated for
CH$_3$NH$_3$PbI$_3$ within DFT (including an estimate for
quasiparticle corrections) we show that band alignments are
consistent with efficient electron transfer to TiO$_2$ and Au 
electrical contacts.


\section{Results}

Optimisation of the crystal structures of NH$_4$PbI$_3$ and
CH$_3$NH$_3$PbI$_3$ have recently been reported\cite{brivio-042111} in DFT using the
PBEsol\cite{pbesol} exchange-correlation functional.  Atomic
forces were converged to within 5 meV/\AA,
and the bond lengths are in good agreement with experiment.  
The representative
$\left<100\right>$ configuration of MA is considered here.
Lattice vectors of these perovskites are approximately cubic
($a$ = 6.29 \AA ~ and 6.21 \AA ~ for the MA and NH$_4$ perovskites, respectively),
with small distortions of the simple cubic ones.  The
valence band maximum and conduction band minimum falls close to a zone
boundary point, the analogue of the $R$ point
$(\frac{1}{2},\frac{1}{2},\frac{1}{2})$ in cubic symmetry. We denote
this point as \textit{R} in the remainder of the paper.

\subsection{Band structure}
The QP band structures for CH$_3$NH$_3$PbI$_3$ and 
NH$_4$PbI$_3$, with colours denoting the orbital character of the
states, are shown in Fig. 1. The ions within the inorganic
(PbI$_3$)$^-$ cage have formal electronic configurations of Pb:
$5d^{10}6s^{2}6p^{0}$ and I: $5p^{6}$. As can be seen from the
colour coding, the valence band maximum consists of approximately
70\% I 5$p$ and 25\% Pb 6$s$ (the Pb 6$s$ forms a band centred
around $-$8\,eV), while the conduction band consists a mixture of
Pb 6$p$ and other orbitals.  The molecular units CH$_3$NH$_3$
and NH$_4$ form $\sigma$ bonds deep in the valence
band.  They are essentially dispersionless: they do not hybridise
with the cage until energies exceed $E_F{+}5$\,eV.  Thus their
interaction with the host is largely electrostatic and structural; 
they provide charge compensation to the PbI$_3^-$ cage.

\begin{figure}[t]
\includegraphics[height=7.5cm]{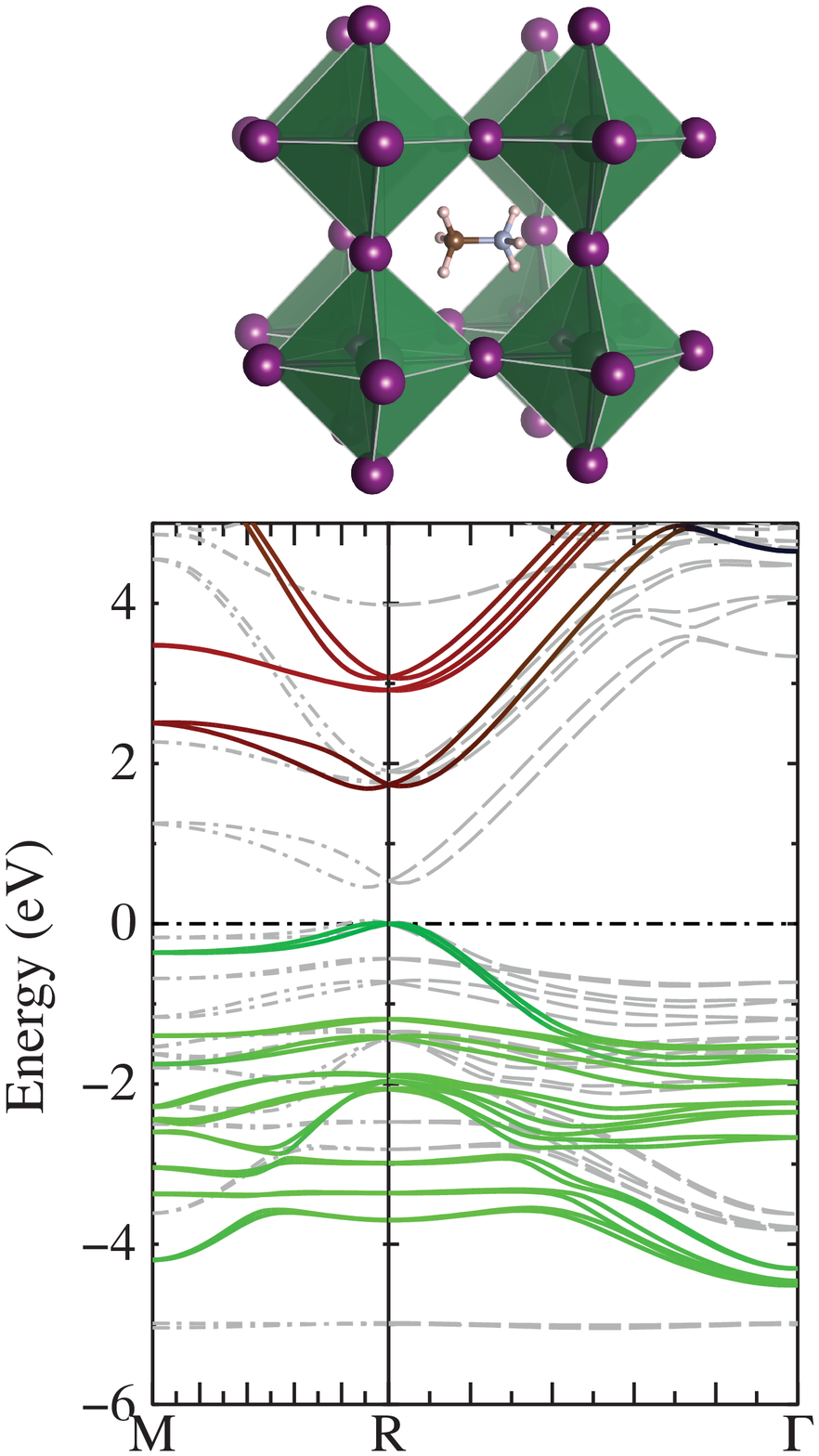} \
\includegraphics[height=7.5cm]{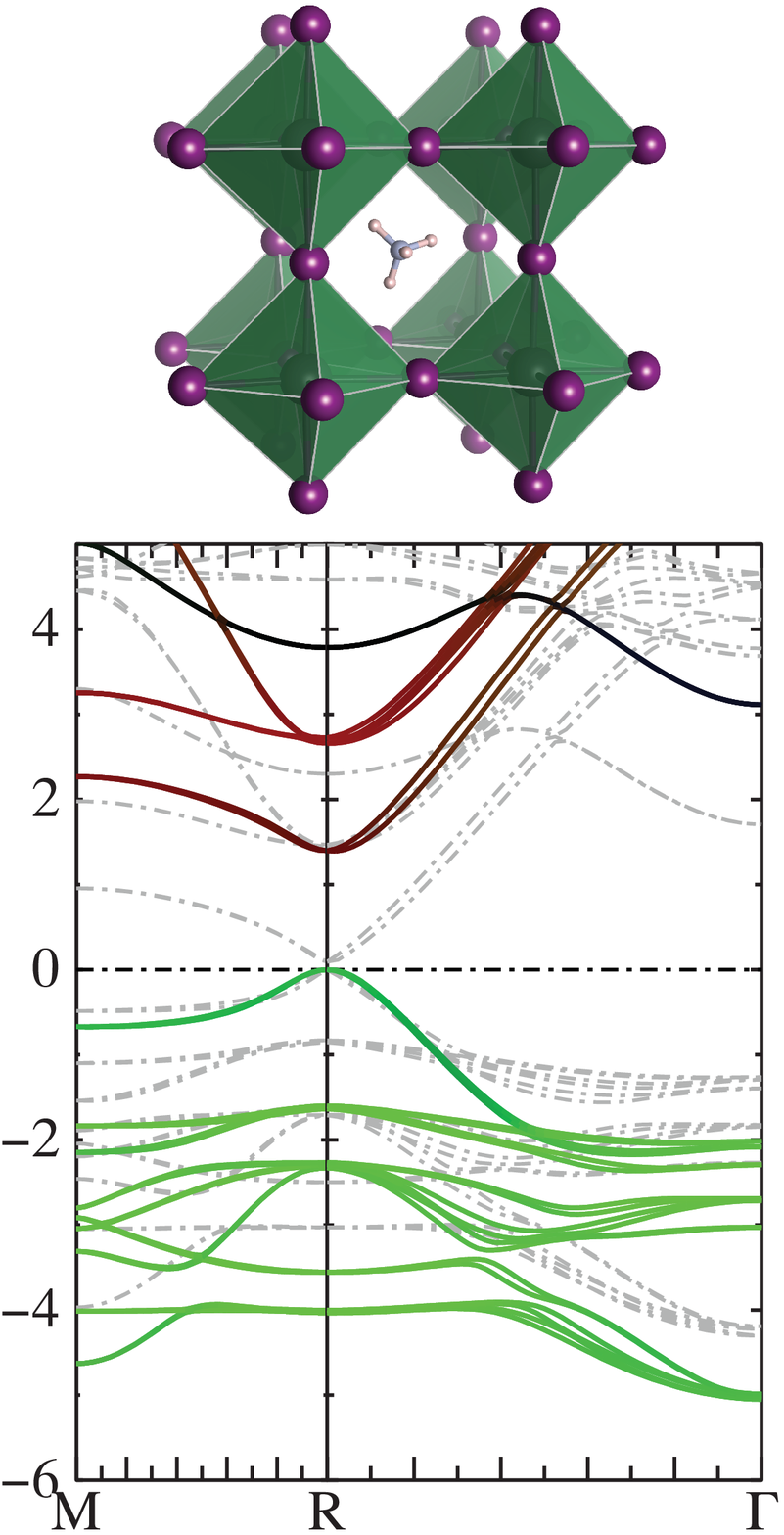}
\caption {\footnotesize \raggedright { {QS\emph{GW} band structure for
      CH$_3$NH$_3$PbI$_3$ (left) and NH$_4$PbI$_3$ (right).  Zero
      denotes valence band maximum.  Bands are
      colored according to their orbital character: green depicts I
      5$p$, red Pb 6$p$, and blue Pb 6$s$.  Points denoted \textit{M}
      and \textit{R} are
      zone-boundary points close to ($\frac{1}{2}$,$\frac{1}{2}$,0)
      and ($\frac{1}{2}$,$\frac{1}{2}$,$\frac{1}{2}$),
      respectively.  The valence band maximum (VBM) and conduction band
      minimum are shifted slightly from \textit{R} as a consequence of the
      $L{\cdot}S$ coupling.  Valence bands near $-2$\,eV
      (conduction bands near +3\,eV) are almost purely green
      (red) showing that they consist largely of I $5p$ (Pb $6p$) character.
      Bands nearer the gap are darker as a result of intermixing with other states.
      Light dashed gray lines show corresponding bands in the LDA.
      The dispersionless state near $-$5\,eV corresponds to a molecular
      level of methylammonium.
      In QS\emph{GW} this state is pushed down to $-$7.7\,eV.
      The dispersion of the highest valence bands is very poorly
      described by the LDA, as described in the text.}
}}
\label{fig:QSGWbands}
\end{figure}

The results presented in Table\,\ref{tab:bandgaps} demonstrate the
various contributions to the band energies around the fundamental gap.
The contribution from SOC ($\sim$1~eV), is extraordinarily large, of
the order of the gap itself; so large that screening is enhanced.  As
a result there is a smaller, but nonetheless non-negligible
contribution of SOC to the electron self-energy ($\Sigma{=}iGW$),
apparent from the difference between `SO($T$)' and `SO($\Sigma$)'.
Furthermore, Table\,\ref{tab:bandgaps} emphasises the importance of
the feedback between $W$ and QP when calculating the band structure of
these systems. The `$GW^{\mathrm{0}}$' gap is based on a perturbation
of the LDA gap; and it is significantly smaller.  Because the LDA gap
is too small, $W$ is overscreened, and $GW$ understimated.  The role of
feedback is important in other semiconductors: it is particularly
strong in InN\cite{Usuda04} and Cu(In,Ga)Se$_2$.  In the latter case
the interplay between $W$ and $E_G$ was shown explicitly by comparing
functionals that did or did not include the dependence of $W$ on band
structure\cite{vidal-056401}.  $W$ and the gap correction is not a
function of the fundamental gap alone: all the bands (including
valence band dispersions) shift in a nontrivial manner.  To reliably
determine the electronic structure including the fundamental gap,
self-consistency is essential.

Recent measurements place the room temperature band gap of
NH$_3$CH$_3$PbI$_3$ at 1.61 eV,\cite{yamada-032302} which falls slightly below the
QS\emph{GW} result.  Some tendency for QS\emph{GW} to
overestimate gaps is expected.  In any case theory and experiment
cannot be compared to better than 0.1 eV resolution for several
reasons.  There are small issues with $k$-point
convergence,\cite{convergence} and with the shape of local wave
functions determined by solving a scalar relativistic equation
rather than the Dirac equation (see the Appendix). On the
experimental side there may be some temperature dependence of the
gap given the structural flexibility of the material;\cite{poglitsch1987dynamic}
this has yet to be explored.

\begin{table}[htbp]
\vbox{\vskip 6pt}
\begin{tabular}{|l|cccc|c|}
\hline
                    & \multicolumn{3}{c}{DFT}                &        & \\
                    & PBEsol & LDA &  +SO            &        & Expt\cite{yamada-032302}  \\
\hline
NH$_3$CH$_3$PbI$_3$ &   1.38  &    1.46     & \ \ 0.53        &        & 1.61 (RT) \\
NH$_4$PbI$_3$       &   1.20  &    1.13     & \ \ 0.09        &        & - \\
\hline
\vbox{\vskip 3pt}
                    &\multicolumn{4}{c}{QS\emph{GW}}                  & \\
                    &   SO=0 &  SO($T$)    & SO($\Sigma$)    & $GW^\mathrm{0}$ & Expt  \\
\hline
NH$_3$CH$_3$PbI$_3$ &   2.73 &  1.78       &   1.67          &  1.27  &  1.61 (RT) \\
NH$_4$PbI$_3$       &   2.30 &  1.36       &   1.38          &  0.76  &   -   \\
\hline
\end{tabular}
  \caption{\footnotesize\raggedright Fundamental band gaps (in eV) of
  CH$_3$NH$_3$PbI$_3$ and NH$_4$PbI$_3$ calculated at varying
  levels of approximation.  
  Top rows show DFT results
  using PBEsol (reported in Ref. \onlinecite{brivio-042111})
   and the Barth-Hedin LDA functional.  First columns
  show that semilocal and local functionals generate similar
  gaps.  $\lambda \mathbf{L}{\cdot}\mathbf{S}$ (column `+SO') strongly reduces
  the gap.  \emph{GW} gaps are shown without spin-orbit coupling (SO=0), with
  $\lambda \mathbf{L}{\cdot}\mathbf{S}$  added to a fixed potential, modifying
  the kinetic energy only (SO($T$)), and included in the
  QS\emph{GW} self-consistency cycle (SO($\Sigma$)).  Column
  `$GW^\mathrm{0}$' is similar to SO($\Sigma$) but $G$ and $W$
  are generated from the LDA.  (In this calculation the full
  $\Sigma$ matrix was generated, not just the diagonal part as is
  customary.  A $Z$ factor of 1 was used to take partial account
  of self-consistency, which brings the gap in better agreement with
  the QS\emph{GW} result; see Appendix A in Ref. \onlinecite{mark06adeq}).
  An error of order 0.1\,eV might be associated with the
  treatment of SOC; see the Appendix.  Room temperature (RT)
  band gap data is only available for NH$_3$CH$_3$PbI$_3$.\cite{yamada-032302}}
\label{tab:bandgaps}
\end{table}

Fig.~\ref{fig:QSGWbands} also shows the LDA energy bands.
Remarkably, the LDA badly underestimates not only the gap but
poorly describes the dispersion in the valence bands.  The
tendency for LDA to underestimate band gaps is traditionally
associated with the energy cost for an excited electron-hole
pair.  The exchange-correlation potential should distinguish
between a neutral excitation (e.g. a hole shifting from one $k$
point in the valence band to another) and one where charge is
separated (excitation of an electron-hole pair).  Such a
distinction is problematic for a local potential, which is by
necessity the same for all electrons.  Such an error is seen in
the present case, as the LDA gap is too small.  As
Fig.~\ref{fig:QSGWbands} clearly shows, LDA and QS\emph{GW}
valence bands also deviate strongly from one another.  Note in
particular the states at \textit{R} between 0 and $-2$\,eV.  
This behaviour allows us to deduce that the hopping matrix 
elements between I $5p$ (and to some
extent Pb $6s$) states are poorly described by the LDA.

\subsection{Carrier effective mass}
Typically light hole masses are too small in the LDA because,
according to $k{\cdot}p$ theory, $m_v^* \propto E_G/V^2$; $V$
is a matrix element of the gradient operator between the
conduction and valence bands. $m_v^*$ is expected to be too small
because $E_G$ is underestimated; indeed for traditional narrow
gap tetrahedral semiconductors, the proportionality between
$m_v^*$ and $E_G$ is reasonably well obeyed.  The LDA predicts
the light hole mass to be too small while other masses (which do
not couple to the nearest conduction band) are reasonably
described.  For example, GaAs has a gap similar to
CH$_3$NH$_3$PbI$_3$, and LDA underestimates it by a comparable
amount ($\sim$1\,eV).  Following expectations, the LDA
underestimates the light hole mass in GaAs by a factor of
$\sim$3.  But for CH$_3$NH$_3$PbI$_3$ the situation is reversed:
the LDA \emph{overestimates} $m_v^*$ even while it severely
underestimates $E_G$.  For NH$_4$PbI$_3$ the LDA and
QS\emph{GW} masses are comparable, but only because the LDA gap
is very small.

\begin{figure}[htbp]
\includegraphics[height=3.5cm]{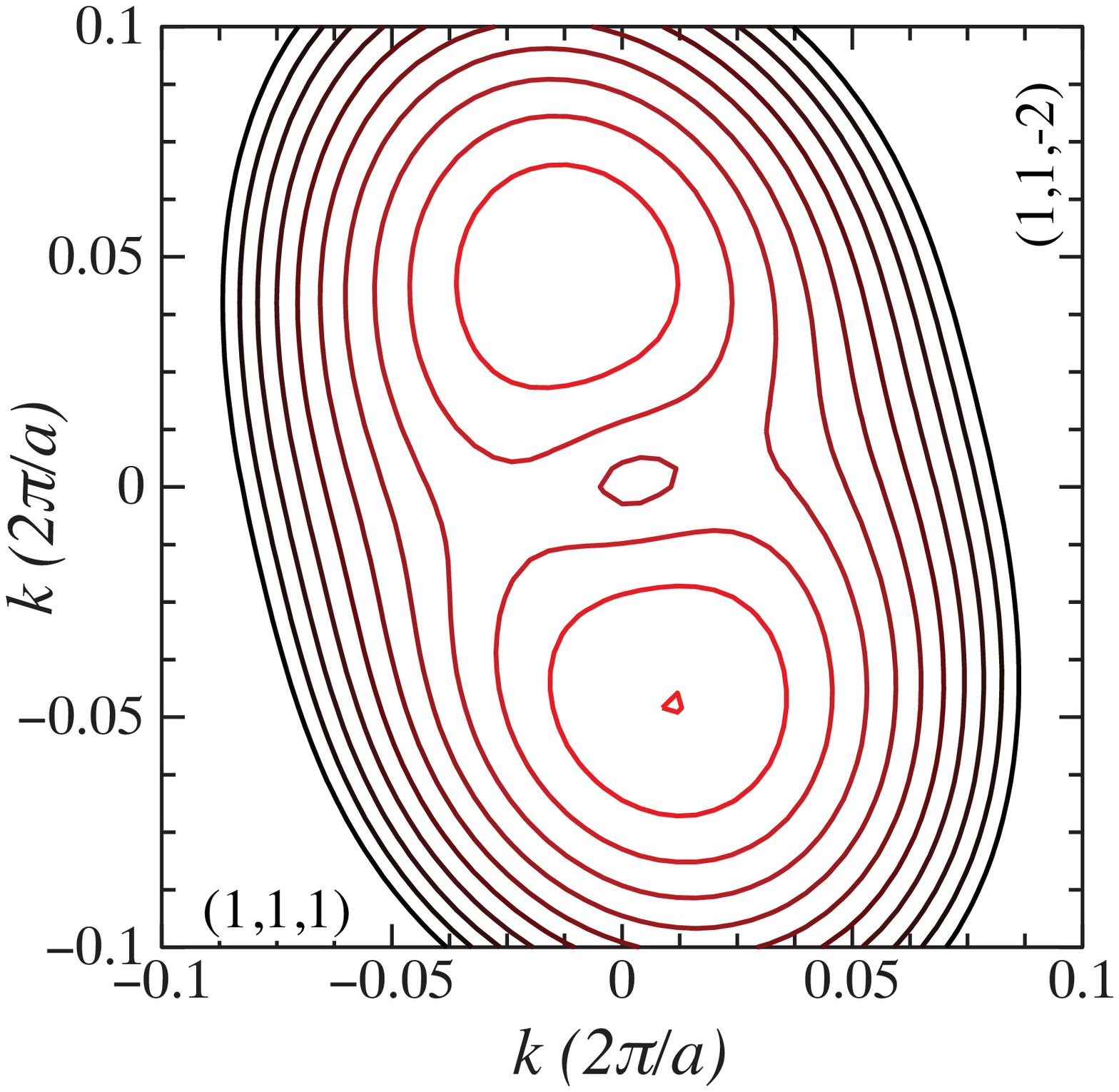}\
\includegraphics[height=3.3cm]{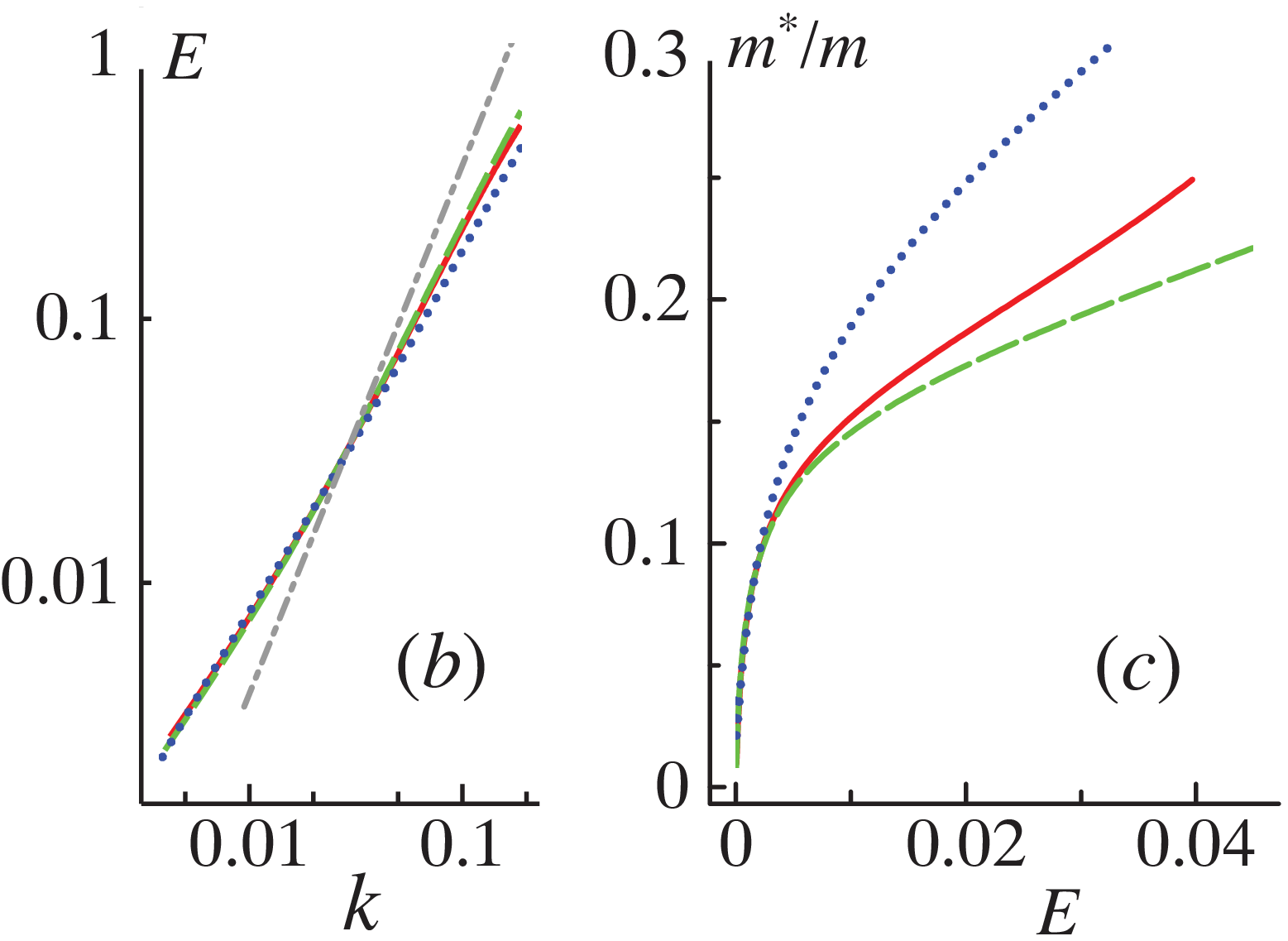}
\caption {\footnotesize \raggedright
{ Left: constant energy contours in the ($1\bar{1}0$) plane 
for the upper valence band of CH$_3$NH$_3$PbI$_3$.
  The origin corresponds to the \textit{R} point and [$111$] and [$11\bar{2}$] are the horizontal and
  vertical axes of $k$.
  Energy contours are in increments of 2.5\,meV,
  so that the outermost contour corresponds
  approximately to RT.  Corresponding contours
  in the ($111$) plane (not shown) appear similar.  Valence bands
  have two maximal points near \textit{R}$\pm$0.005[$11\bar{2}$]. At
  low temperature and low doping, ($E_F{<}5$\,meV) the two extrema
  act as independent centres with approximately spherical
  effective masses.  At high doping ($E_F{>}20$\,meV) or high
  temperature, holes effectively see a single band maximum with
  roughly elliptical constant energy surfaces.
  Panels ($b$) and ($c$) show energy $E_v(k){\equiv}E(R){-}E(k)$ for the
  lower valence band.  This band has a single maximum at $R$, with approximately
  spherical dispersion.  Panel $(b)$ shows $E_v(k)$, on a
  log-log scale in the [$1\bar{1}0$], [$111$], and [$11\bar{2}$] directions as red
  solid, green dashed and blue dotted lines, respectively.
  For comparison a parabolic band with effective mass 0.1$m$ is shown as a grey dot-dashed line.
  Panel ($c$) plots $h^2k^2/(2m\,E_v)$ against $E_v$, which may
  be taken as a definition of the effective mass (see text).
  $E_v$ is in eV, $k$ in units $2\pi/a$.}
}
\label{fig:masses}
\end{figure}

Spin-orbit coupling greatly complicates both valence and
conduction bands within $k_{B}T$ of the band edges.  We
focus on the two valence bands of NH$_3$CH$_3$PbI$_3$, as these
are the ones that govern transport in hole-based devices.  As
these two bands approach the \textit{R} point, they must merge to the same
value by symmetry. However, they approach the \textit{R} point with a
linear dispersion in some directions; as a consequence the
dispersions in the upper and lower bands are non-analytic.  The upper band is
maximum in some directions but increases with a linear slope in
the $\pm$[$11\bar{2}$] direction. Two maxima form near
\textit{R}$\pm$0.005$\times$[$11\bar{2}$], Fig.~\ref{fig:masses}($a$),
arising from Dresselhaus spin-orbit coupling, which is
even more pronounced in the lower conduction band.
The lower band has a single maximum at \textit{R}, and its constant energy
surfaces deviate only modestly from spheres for $k$ near \emph{R}
(Fig.~\ref{fig:masses}($b$)).  Yet, the right panels of
Fig.~\ref{fig:masses} show $E_v(k){\equiv}E(R){-}E(k)$ deviates markedly from a
parabolic dispersion.  This has important consequences for the
device behavior of this material.  Fig.~\ref{fig:masses}($c$)
shows that, provided $E_v{>}10$\,meV, the band dispersion
can be expressed approximately as a $k$ dependent mass:
\begin{equation}
\frac{h^2k^2}{2m} = m^*(k)E_v(k),\quad
\frac{m^*(k)}{m} = m_0 [ 1 + \alpha E_v(k)]
\end{equation}
with $m_0^*$ $\sim$ 0.12 and $\alpha$ independent of $|\mathbf{k}|$
but dependent on orientation.  For $E_v$ large enough, the upper
valence band can also be described by an effective mass of
roughly the same size.  
The lower conduction band exhibits a similar behaviour, with $m_0^*$ $\sim$ 0.15.
The small masses explain how these
materials can exhibit high mobility and long diffusion lengths.
Bands of NH$_4$PbI$_3$ differ in important details from
NH$_3$CH$_3$PbI$_3$ (the influence of SOC is less pronounced),
but the basic structure is similar.  Both the band gap and
effective masses are reduced relative to NH$_3$CH$_3$PbI$_3$, as
can be seen directly by inspecting Fig.~\ref{fig:QSGWbands}.

\subsection{Optical and dielectric response}
\begin{figure}[htbp]
\includegraphics[height=4.5cm]{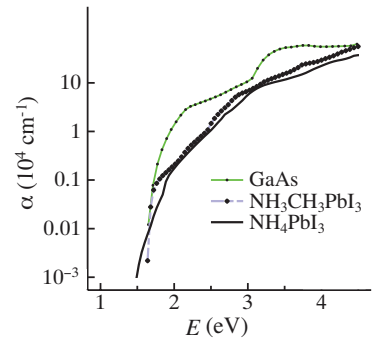}
\caption {\footnotesize \raggedright { {Optical absorption spectrum calculated
      within the RPA from the QS\emph{GW} potential, for
      CH$_3$NH$_3$PbI$_3$ and NH$_4$PbI$_3$. The absorption
      is smaller than, but comparable to that of GaAs,
      shown for comparison. Note that similar measurements 
      have been reported in Ref. \onlinecite{wolf-2014}.
      }  }}
\label{fig:QSGWalpha}
\end{figure}

It is known that these compounds strongly absorb visible light.  We
confirm this through a random phase approximation (RPA) calculation of $\alpha(\omega)$ 
from the imaginary part of the macroscopic dielectric function
$\epsilon_M(\omega) =
[\epsilon^{-1}_{\mathbf{G},\mathbf{G}^\prime=0})(\mathbf{q}{\to}0,\omega)]^{-1}$.
As Fig.~\ref{fig:QSGWalpha} shows, $\alpha$ is somewhat smaller than -- but
comparable to -- that of GaAs. This explains why very thin layers of the hybrid perovskites
have been found to give high photovoltaic efficiencies.
Indeed combined with the low carrier effective masses, the resulting
electron-hole diffusion lengths exceed the typical film thickness.

\begin{table}[htbp]
\vbox{\vskip 6pt}
\begin{tabular}{|l|ccc|cc|}
\hline
                    & \multicolumn{3}{c|}{PBEsol}                      &  \multicolumn{2}{c|}{QS\emph{GW}} \\
                    &  $E_G$   & $\epsilon_{0}$ & $\epsilon_{\infty}$ &  $E_G$  & $\epsilon_{\infty}$ \\
\hline
NH$_3$CH$_3$PbI$_3$ &   1.38  &    25.7     & 6.1                      &  1.67     & 4.5 \\
NH$_4$PbI$_3$       &   1.20  &    18.4     & 6.5                      &  1.38     & 5.0 \\
\hline
\end{tabular}
\caption{\footnotesize\raggedright 
Dielectric constants (isotropic average of the tensor) and band gaps (eV), calculated in density-functional perturbation theory without SOC (from Ref. \onlinecite{brivio-042111}), and in the RPA with SOC.}
\label{tab:dielectric}
\end{table}

Some static ($\epsilon_{0}$) and high-frequency ($\epsilon_{\infty}$)
dielectric constants are shown in Table\,\ref{tab:dielectric}.
These values were unusually sensitive to the \textit{k}-point sampling density
and require dense meshes for convergence.
Those calculated by density-functional
perturbation theory (e.g. PBEsol) leave out SOC and thus get fortuitously
good band gaps.  As a result $\epsilon_{\infty}$ is not so different from
the QS\emph{GW} case (which includes SOC).  Contributions to
$\epsilon_{0}$ from lattice polarization are significantly larger than
seen in typical tetrahedral semiconductors (compare $\epsilon_{0}$ to
$\epsilon_{\infty}$).
For NH$_3$CH$_3$PbI$_3$, the average value of the static dielectric 
constant, including quasi-particle (QP) corrections from Table \ref{tab:dielectric}, of 24.1
is in very good agreement with permittivity measurements of 23.3.\cite{onoda-935}
These values exclude contributions from orientational 
disorder of the methylammonium ions, 
which is the subject of further study.\cite{frost-2014}

\subsection{Surface ionisation potential}
In order to place the electronic bands on an absolute energy scale, we
have aligned the quasiparticle energies with respect to the
vacuum level of a non-polar (110) termination of the perovskite,
generated using \textit{METADISE}\cite{metadise}. 
We take the Pb 1$s$ core level as an energy reference and 
use a planar average of the electrostatic potential following 
the standard procedure.\cite{butler-115320,butler-2014} 
 The slab
model consisted of four perovskite layers with the dipole of the methyl
ammonium cations aligned parallel to the surface termination, 
which ensure no macroscopic electric field. 
The
resulting ionisation potential is 5.7 eV 
(5.9 eV from LDA, which is corrected by the \emph{GW}$^0$
self energy), with a corresponding electron affinity of 4.0 eV. 
These values are in good agreement with
initial photoemission measurements of thin films (5.4 eV) and explain the success
of TiO$_2$ (electron) and Au (hole) contacts.\cite{park-2423} 

\section{Conclusion}
We have explored the electronic structure of
two key hybrid halide perovskites. Relativistic and many-body corrections
are shown to be essential for a quantitative description
of the bulk properties important for photovoltaics: 
band gap, band dispersion,
effective mass and dielectric response. These organic-inorganic
materials display quantum mechanical behaviour atypical of traditional
semiconductors, which begins to explain their remarkable performance in
mesoporous and thin-film solar cells.

\acknowledgments
We thank L.M. Peter and H. J Snaith for useful discussions, and acknowledge membership of the UK's HPC Materials Chemistry Consortium, which is funded by EPSRC grant EP/F067496. F.B., K.T.B and A.W. are funded by the DESTINY ITN (316494), EPSRC (EP/J017361/1) and the ERC (277757), respectively.

\section*{Appendix: Simplified treatment of Spin-Orbit Coupling in QS\emph{GW}}
Aryasetiwan and Biermann \cite{Aryasetiawan08} developed a formalism
for \emph{GW} with spin-dependent interactions.  Rather than proceed
with a completely non-collinear treatment, we take advantage of the
fact that $\lambda \mathbf{L}{\cdot}\mathbf{S}$ is relatively small,
and moreover that the non-collinear part of the eigenfunctions is
unimportant for these semiconductors.  We present a simplified
treatment that generates results essentially as good as adding
$\lambda \mathbf{L}{\cdot}\mathbf{S}$ non-perturbatively to the scalar
Dirac Hamiltonian for M-PbI$_{3}$ compounds.

Partitioning $\mathbf{L}{\cdot}\mathbf{S}$ into components,
the non-interacting QS\emph{GW} Hamiltonian reads
\[ H_0 = H_0(\lambda{=}0) + {\lambda}L^zS^z + {\lambda}(\mathbf{L}^+{\cdot}\mathbf{S}^- + \mathbf{L}^-{\cdot}\mathbf{S}^+) \]
The first two terms are spin-diagonal and can be diagonalized
non-perturbatively in the same manner as $H_0(\lambda{=}0)$.  The
eigenvalues $\epsilon_i$ and eigenfunctions $\psi_i$ contain the
$L^zS^z$ portion of $\mathbf{L}{\cdot}\mathbf{S}$, keeping $\psi_i$
spin diagonal.  The latter two terms, when treated exactly, further
shift the $\epsilon_i$ and also introduce spin off-diagonal parts to
the $\psi_i$.  We allow the former but omit the latter.

The lowest order of correction to the eigenvalues is second order and
we follow the spirit of second-order perturbation theory.  Let
$\delta_{ij}$ be the initial splitting in $\epsilon_i$ and
$\epsilon_j$, $\Delta_{ij}{=}|\epsilon_i{-}\epsilon_j|/2$. If
$H^{+-}_{ij}$ couples $i$ and $j$,
$\Delta_{ij}$ increases by $ \delta \Delta_{ij} =
|H^{+-}_{ij}|^2/|\epsilon_{i} -\epsilon_{j}|$, in lowest order.

Second-order perturbation theory can be problematic when $\epsilon_i{\to}\epsilon_j$.
We instead obtain $\delta \Delta_{ij}$ from
\[ \delta \Delta_{ij} = \sqrt{\Delta_{ij}^2  + |H^{+-}_{ij}|^2 } - |\Delta_{ij}| \]
This expression is exact if $i$ and $j$ are isolated from the rest of
the system.  The final expression (the net shift for each $\epsilon_i$
is obtained by summing over each $ij$ pair) is nevertheless correct
only to second order because terms involving three or more states are
not included.

We carefully tested our quasi-perturbative approach in the LDA or LDA+$U$
context for a wide range of materials, e.g. Fe, Sn, Au, GdN, Pu, and
the perovskites addressed in this manuscript.  In all cases except Pu
($Z$=94) the difference between the perturbation expression resulted
in $\epsilon_i$ very close to $\mathbf{L}{\cdot}\mathbf{S}$ treated
non-perturbatively.  For CH$_3$NH$_3$PbI$_3$, for example, $E_G$
changed by less than 0.01\,eV.  Self-consistency carried through with
both approaches generate a slight difference in density, but no
significant difference in the $\epsilon_i$.

Tests of the adequacy of the quasi-perturbative
$\mathbf{L}{\cdot}\mathbf{S}$ in the QS\emph{GW} were performed as
follows: self-consistency was reached with
$\mathbf{L}{\cdot}\mathbf{S}$ included quasi-perturbatively, and for a
given $\Sigma$, the quasiparticle levels with
$\mathbf{L}{\cdot}\mathbf{S}$ calculated non-perturbatively were
compared to the perturbative treatment.  As in the LDA case, negligible
differences were found for all compounds studied except for Pu, where
modest differences were found.
As in the LDA case, the
nonperturbative treatment generated a slight change in density.  Since
the $\epsilon_i$ are reliably determined, it is unlikely that a better
treatment of $\mathbf{L}{\cdot}\mathbf{S}$ (non-collinear
eigenfunctions) will further affect $\Sigma$ appreciably in these
compounds.  On the other hand, fully relativistic treatment might
affect $H_0$ a little, since the relativistic radial functions vary as
$r^{\gamma}$ for small $r$, where $\gamma^2 = \kappa^2 {-} (2Z/c)^2$,
$\kappa$ playing the role of the $l$ quantum number.  $\gamma$ reduces
the scalar relativistic case only when $c{\to}\infty$.  A better
treatment of the small-$r$ behavior of the partial waves modifies spin
orbit splitting of the $p$ levels for Pb by about 10\%, which is not
included here.

An appreciable effect of $\mathbf{L}{\cdot}\mathbf{S}$ on $\Sigma$ is
observed only for compounds with large-$Z$ constituents.  For
semiconductors as heavy as Sn ($Z$=50), and for metals as heavy as Au
($Z$=79), the effect of $\mathbf{L}{\cdot}\mathbf{S}$ on $\Sigma$
appears to be very small.  But for the iodide perovskites studied
here, $\mathbf{L}{\cdot}\mathbf{S}$ has a noticeable effect on $\Sigma$
(Table\,\ref{tab:bandgaps}) because of the interplay between $E_G$ and
$\epsilon$ present in semiconductors but not in metals.

\bibliography{library}

\begin{thebibliography}{38}%
\makeatletter
\providecommand \@ifxundefined [1]{%
 \@ifx{#1\undefined}
}%
\providecommand \@ifnum [1]{%
 \ifnum #1\expandafter \@firstoftwo
 \else \expandafter \@secondoftwo
 \fi
}%
\providecommand \@ifx [1]{%
 \ifx #1\expandafter \@firstoftwo
 \else \expandafter \@secondoftwo
 \fi
}%
\providecommand \natexlab [1]{#1}%
\providecommand \enquote  [1]{``#1''}%
\providecommand \bibnamefont  [1]{#1}%
\providecommand \bibfnamefont [1]{#1}%
\providecommand \citenamefont [1]{#1}%
\providecommand \href@noop [0]{\@secondoftwo}%
\providecommand \href [0]{\begingroup \@sanitize@url \@href}%
\providecommand \@href[1]{\@@startlink{#1}\@@href}%
\providecommand \@@href[1]{\endgroup#1\@@endlink}%
\providecommand \@sanitize@url [0]{\catcode `\\12\catcode `\$12\catcode
  `\&12\catcode `\#12\catcode `\^12\catcode `\_12\catcode `\%12\relax}%
\providecommand \@@startlink[1]{}%
\providecommand \@@endlink[0]{}%
\providecommand \url  [0]{\begingroup\@sanitize@url \@url }%
\providecommand \@url [1]{\endgroup\@href {#1}{\urlprefix }}%
\providecommand \urlprefix  [0]{URL }%
\providecommand \Eprint [0]{\href }%
\providecommand \doibase [0]{http://dx.doi.org/}%
\providecommand \selectlanguage [0]{\@gobble}%
\providecommand \bibinfo  [0]{\@secondoftwo}%
\providecommand \bibfield  [0]{\@secondoftwo}%
\providecommand \translation [1]{[#1]}%
\providecommand \BibitemOpen [0]{}%
\providecommand \bibitemStop [0]{}%
\providecommand \bibitemNoStop [0]{.\EOS\space}%
\providecommand \EOS [0]{\spacefactor3000\relax}%
\providecommand \BibitemShut  [1]{\csname bibitem#1\endcsname}%
\let\auto@bib@innerbib\@empty
\bibitem [{\citenamefont {Kojima}\ \emph {et~al.}(2009)\citenamefont {Kojima},
  \citenamefont {Teshima}, \citenamefont {Shirai},\ and\ \citenamefont
  {Miyasaka}}]{kojima-6050}%
  \BibitemOpen
  \bibfield  {author} {\bibinfo {author} {\bibfnamefont {A.}~\bibnamefont
  {Kojima}}, \bibinfo {author} {\bibfnamefont {K.}~\bibnamefont {Teshima}},
  \bibinfo {author} {\bibfnamefont {Y.}~\bibnamefont {Shirai}}, \ and\ \bibinfo
  {author} {\bibfnamefont {T.}~\bibnamefont {Miyasaka}},\ }\href@noop {}
  {\bibfield  {journal} {\bibinfo  {journal} {J. Am. Chem. Soc.}\ }\textbf
  {\bibinfo {volume} {131}},\ \bibinfo {pages} {6050} (\bibinfo {year}
  {2009})}\BibitemShut {NoStop}%
\bibitem [{\citenamefont {Lee}\ \emph {et~al.}(2012)\citenamefont {Lee},
  \citenamefont {Teuscher}, \citenamefont {Miyasaka}, \citenamefont
  {Murakami},\ and\ \citenamefont {Snaith}}]{snaith-1}%
  \BibitemOpen
  \bibfield  {author} {\bibinfo {author} {\bibfnamefont {M.~M.}\ \bibnamefont
  {Lee}}, \bibinfo {author} {\bibfnamefont {J.}~\bibnamefont {Teuscher}},
  \bibinfo {author} {\bibfnamefont {T.}~\bibnamefont {Miyasaka}}, \bibinfo
  {author} {\bibfnamefont {T.~N.}\ \bibnamefont {Murakami}}, \ and\ \bibinfo
  {author} {\bibfnamefont {H.~J.}\ \bibnamefont {Snaith}},\ }\href@noop {}
  {\bibfield  {journal} {\bibinfo  {journal} {Science}\ }\textbf {\bibinfo
  {volume} {338}},\ \bibinfo {pages} {643} (\bibinfo {year}
  {2012})}\BibitemShut {NoStop}%
\bibitem [{\citenamefont {Burschka}\ \emph {et~al.}(2013)\citenamefont
  {Burschka}, \citenamefont {Pellet}, \citenamefont {Moon}, \citenamefont
  {Humphry-Baker}, \citenamefont {Gao}, \citenamefont {Nazeeruddin},\ and\
  \citenamefont {Gr{\"a}tzel}}]{gratzel-1}%
  \BibitemOpen
  \bibfield  {author} {\bibinfo {author} {\bibfnamefont {J.}~\bibnamefont
  {Burschka}}, \bibinfo {author} {\bibfnamefont {N.}~\bibnamefont {Pellet}},
  \bibinfo {author} {\bibfnamefont {S.-J.}\ \bibnamefont {Moon}}, \bibinfo
  {author} {\bibfnamefont {R.}~\bibnamefont {Humphry-Baker}}, \bibinfo {author}
  {\bibfnamefont {P.}~\bibnamefont {Gao}}, \bibinfo {author} {\bibfnamefont
  {M.~K.}\ \bibnamefont {Nazeeruddin}}, \ and\ \bibinfo {author} {\bibfnamefont
  {M.}~\bibnamefont {Gr{\"a}tzel}},\ }\href@noop {} {\bibfield  {journal}
  {\bibinfo  {journal} {Nature}\ }\textbf {\bibinfo {volume} {499}},\ \bibinfo
  {pages} {316} (\bibinfo {year} {2013})}\BibitemShut {NoStop}%
\bibitem [{\citenamefont {Kim}\ \emph {et~al.}(2012)\citenamefont {Kim},
  \citenamefont {Lee}, \citenamefont {Im}, \citenamefont {Lee}, \citenamefont
  {Moehl}, \citenamefont {Marchioro}, \citenamefont {Moon}, \citenamefont
  {Humphry-Baker}, \citenamefont {Yum},\ and\ \citenamefont
  {Moser}}]{gratzel-2}%
  \BibitemOpen
  \bibfield  {author} {\bibinfo {author} {\bibfnamefont {H.-S.}\ \bibnamefont
  {Kim}}, \bibinfo {author} {\bibfnamefont {C.-R.}\ \bibnamefont {Lee}},
  \bibinfo {author} {\bibfnamefont {J.-H.}\ \bibnamefont {Im}}, \bibinfo
  {author} {\bibfnamefont {K.-B.}\ \bibnamefont {Lee}}, \bibinfo {author}
  {\bibfnamefont {T.}~\bibnamefont {Moehl}}, \bibinfo {author} {\bibfnamefont
  {A.}~\bibnamefont {Marchioro}}, \bibinfo {author} {\bibfnamefont {S.-J.}\
  \bibnamefont {Moon}}, \bibinfo {author} {\bibfnamefont {R.}~\bibnamefont
  {Humphry-Baker}}, \bibinfo {author} {\bibfnamefont {J.-H.}\ \bibnamefont
  {Yum}}, \ and\ \bibinfo {author} {\bibfnamefont {J.~E.}\ \bibnamefont
  {Moser}},\ }\href@noop {} {\bibfield  {journal} {\bibinfo  {journal} {Sci.
  Rep.}\ }\textbf {\bibinfo {volume} {2}},\ \bibinfo {pages} {591} (\bibinfo
  {year} {2012})}\BibitemShut {NoStop}%
\bibitem [{\citenamefont {Heo}\ \emph {et~al.}(2013)\citenamefont {Heo},
  \citenamefont {Im}, \citenamefont {Noh}, \citenamefont {Mandal},
  \citenamefont {Lim}, \citenamefont {Chang}, \citenamefont {Lee},
  \citenamefont {Kim}, \citenamefont {Sarkar},\ and\ \citenamefont
  {Nazeeruddin}}]{heo-486}%
  \BibitemOpen
  \bibfield  {author} {\bibinfo {author} {\bibfnamefont {J.~H.}\ \bibnamefont
  {Heo}}, \bibinfo {author} {\bibfnamefont {S.~H.}\ \bibnamefont {Im}},
  \bibinfo {author} {\bibfnamefont {J.~H.}\ \bibnamefont {Noh}}, \bibinfo
  {author} {\bibfnamefont {T.~N.}\ \bibnamefont {Mandal}}, \bibinfo {author}
  {\bibfnamefont {C.-S.}\ \bibnamefont {Lim}}, \bibinfo {author} {\bibfnamefont
  {J.~A.}\ \bibnamefont {Chang}}, \bibinfo {author} {\bibfnamefont {Y.~H.}\
  \bibnamefont {Lee}}, \bibinfo {author} {\bibfnamefont {H.-j.}\ \bibnamefont
  {Kim}}, \bibinfo {author} {\bibfnamefont {A.}~\bibnamefont {Sarkar}}, \ and\
  \bibinfo {author} {\bibfnamefont {M.~K.}\ \bibnamefont {Nazeeruddin}},\
  }\href@noop {} {\bibfield  {journal} {\bibinfo  {journal} {Nature Photon.}\
  }\textbf {\bibinfo {volume} {7}},\ \bibinfo {pages} {486} (\bibinfo {year}
  {2013})}\BibitemShut {NoStop}%
\bibitem [{\citenamefont {Carnie}\ \emph {et~al.}(2013)\citenamefont {Carnie},
  \citenamefont {Charbonnaeu}, \citenamefont {Davies}, \citenamefont
  {Troughton}, \citenamefont {Watson}, \citenamefont {Wojciechowski},
  \citenamefont {Snaith},\ and\ \citenamefont {Worsley}}]{snaith-2}%
  \BibitemOpen
  \bibfield  {author} {\bibinfo {author} {\bibfnamefont {M.~J.}\ \bibnamefont
  {Carnie}}, \bibinfo {author} {\bibfnamefont {C.}~\bibnamefont {Charbonnaeu}},
  \bibinfo {author} {\bibfnamefont {M.~L.}\ \bibnamefont {Davies}}, \bibinfo
  {author} {\bibfnamefont {J.}~\bibnamefont {Troughton}}, \bibinfo {author}
  {\bibfnamefont {T.~M.}\ \bibnamefont {Watson}}, \bibinfo {author}
  {\bibfnamefont {K.}~\bibnamefont {Wojciechowski}}, \bibinfo {author}
  {\bibfnamefont {H.}~\bibnamefont {Snaith}}, \ and\ \bibinfo {author}
  {\bibfnamefont {D.~A.}\ \bibnamefont {Worsley}},\ }\href@noop {} {\bibfield
  {journal} {\bibinfo  {journal} {Chem. Commun.}\ }\textbf {\bibinfo {volume}
  {49}},\ \bibinfo {pages} {7893} (\bibinfo {year} {2013})}\BibitemShut
  {NoStop}%
\bibitem [{\citenamefont {Stranks}\ \emph {et~al.}(2013)\citenamefont
  {Stranks}, \citenamefont {Eperon}, \citenamefont {Grancini}, \citenamefont
  {Menelaou}, \citenamefont {Alcocer}, \citenamefont {Leijtens}, \citenamefont
  {Herz}, \citenamefont {Petrozza},\ and\ \citenamefont
  {Snaith}}]{snaith-6156}%
  \BibitemOpen
  \bibfield  {author} {\bibinfo {author} {\bibfnamefont {S.~D.}\ \bibnamefont
  {Stranks}}, \bibinfo {author} {\bibfnamefont {G.~E.}\ \bibnamefont {Eperon}},
  \bibinfo {author} {\bibfnamefont {G.}~\bibnamefont {Grancini}}, \bibinfo
  {author} {\bibfnamefont {C.}~\bibnamefont {Menelaou}}, \bibinfo {author}
  {\bibfnamefont {M.~J.}\ \bibnamefont {Alcocer}}, \bibinfo {author}
  {\bibfnamefont {T.}~\bibnamefont {Leijtens}}, \bibinfo {author}
  {\bibfnamefont {L.~M.}\ \bibnamefont {Herz}}, \bibinfo {author}
  {\bibfnamefont {A.}~\bibnamefont {Petrozza}}, \ and\ \bibinfo {author}
  {\bibfnamefont {H.~J.}\ \bibnamefont {Snaith}},\ }\href@noop {} {\bibfield
  {journal} {\bibinfo  {journal} {Science}\ }\textbf {\bibinfo {volume}
  {342}},\ \bibinfo {pages} {341} (\bibinfo {year} {2013})}\BibitemShut
  {NoStop}%
\bibitem [{\citenamefont {Liu}, \citenamefont {Johnston},\ and\ \citenamefont
  {Snaith}(2013)}]{snaith-7467}%
  \BibitemOpen
  \bibfield  {author} {\bibinfo {author} {\bibfnamefont {M.}~\bibnamefont
  {Liu}}, \bibinfo {author} {\bibfnamefont {M.~B.}\ \bibnamefont {Johnston}}, \
  and\ \bibinfo {author} {\bibfnamefont {H.~J.}\ \bibnamefont {Snaith}},\
  }\href@noop {} {\bibfield  {journal} {\bibinfo  {journal} {Nature}\ }\textbf
  {\bibinfo {volume} {501}},\ \bibinfo {pages} {395} (\bibinfo {year}
  {2013})}\BibitemShut {NoStop}%
\bibitem [{\citenamefont {Xing}\ \emph {et~al.}(2013)\citenamefont {Xing},
  \citenamefont {Mathews}, \citenamefont {Sun}, \citenamefont {Lim},
  \citenamefont {Lam}, \citenamefont {Gr{\"a}tzel}, \citenamefont
  {Mhaisalkar},\ and\ \citenamefont {Sum}}]{gratzel-6156}%
  \BibitemOpen
  \bibfield  {author} {\bibinfo {author} {\bibfnamefont {G.}~\bibnamefont
  {Xing}}, \bibinfo {author} {\bibfnamefont {N.}~\bibnamefont {Mathews}},
  \bibinfo {author} {\bibfnamefont {S.}~\bibnamefont {Sun}}, \bibinfo {author}
  {\bibfnamefont {S.~S.}\ \bibnamefont {Lim}}, \bibinfo {author} {\bibfnamefont
  {Y.~M.}\ \bibnamefont {Lam}}, \bibinfo {author} {\bibfnamefont
  {M.}~\bibnamefont {Gr{\"a}tzel}}, \bibinfo {author} {\bibfnamefont
  {S.}~\bibnamefont {Mhaisalkar}}, \ and\ \bibinfo {author} {\bibfnamefont
  {T.~C.}\ \bibnamefont {Sum}},\ }\href@noop {} {\bibfield  {journal} {\bibinfo
   {journal} {Science}\ }\textbf {\bibinfo {volume} {342}},\ \bibinfo {pages}
  {344} (\bibinfo {year} {2013})}\BibitemShut {NoStop}%
\bibitem [{\citenamefont {Kim}\ \emph {et~al.}(2013)\citenamefont {Kim},
  \citenamefont {Mora-Sero}, \citenamefont {Gonzalez-Pedro}, \citenamefont
  {Fabregat-Santiago}, \citenamefont {Juarez-Perez}, \citenamefont {Park},\
  and\ \citenamefont {Bisquert}}]{bisquert-4}%
  \BibitemOpen
  \bibfield  {author} {\bibinfo {author} {\bibfnamefont {H.-S.}\ \bibnamefont
  {Kim}}, \bibinfo {author} {\bibfnamefont {I.}~\bibnamefont {Mora-Sero}},
  \bibinfo {author} {\bibfnamefont {V.}~\bibnamefont {Gonzalez-Pedro}},
  \bibinfo {author} {\bibfnamefont {F.}~\bibnamefont {Fabregat-Santiago}},
  \bibinfo {author} {\bibfnamefont {E.~J.}\ \bibnamefont {Juarez-Perez}},
  \bibinfo {author} {\bibfnamefont {N.-G.}\ \bibnamefont {Park}}, \ and\
  \bibinfo {author} {\bibfnamefont {J.}~\bibnamefont {Bisquert}},\ }\href@noop
  {} {\bibfield  {journal} {\bibinfo  {journal} {Nature Commun.}\ }\textbf
  {\bibinfo {volume} {4}},\ \bibinfo {pages} {2242} (\bibinfo {year}
  {2013})}\BibitemShut {NoStop}%
\bibitem [{\citenamefont {Baikie}\ \emph {et~al.}(2013)\citenamefont {Baikie},
  \citenamefont {Fang}, \citenamefont {Kadro}, \citenamefont {Schreyer},
  \citenamefont {Wei}, \citenamefont {Mhaisalkar}, \citenamefont {Graetzel},\
  and\ \citenamefont {White}}]{baikie-5628}%
  \BibitemOpen
  \bibfield  {author} {\bibinfo {author} {\bibfnamefont {T.}~\bibnamefont
  {Baikie}}, \bibinfo {author} {\bibfnamefont {Y.}~\bibnamefont {Fang}},
  \bibinfo {author} {\bibfnamefont {J.~M.}\ \bibnamefont {Kadro}}, \bibinfo
  {author} {\bibfnamefont {M.}~\bibnamefont {Schreyer}}, \bibinfo {author}
  {\bibfnamefont {F.}~\bibnamefont {Wei}}, \bibinfo {author} {\bibfnamefont
  {S.~G.}\ \bibnamefont {Mhaisalkar}}, \bibinfo {author} {\bibfnamefont
  {M.}~\bibnamefont {Graetzel}}, \ and\ \bibinfo {author} {\bibfnamefont
  {T.~J.}\ \bibnamefont {White}},\ }\href@noop {} {\bibfield  {journal}
  {\bibinfo  {journal} {J. Mater. Chem. A}\ }\textbf {\bibinfo {volume} {1}},\
  \bibinfo {pages} {5628} (\bibinfo {year} {2013})}\BibitemShut {NoStop}%
\bibitem [{\citenamefont {Calabrese}\ \emph {et~al.}(1991)\citenamefont
  {Calabrese}, \citenamefont {Jones}, \citenamefont {Harlow}, \citenamefont
  {Herron}, \citenamefont {Thorn},\ and\ \citenamefont
  {Wang}}]{calabrese-2328}%
  \BibitemOpen
  \bibfield  {author} {\bibinfo {author} {\bibfnamefont {J.}~\bibnamefont
  {Calabrese}}, \bibinfo {author} {\bibfnamefont {N.}~\bibnamefont {Jones}},
  \bibinfo {author} {\bibfnamefont {R.}~\bibnamefont {Harlow}}, \bibinfo
  {author} {\bibfnamefont {N.}~\bibnamefont {Herron}}, \bibinfo {author}
  {\bibfnamefont {D.}~\bibnamefont {Thorn}}, \ and\ \bibinfo {author}
  {\bibfnamefont {Y.}~\bibnamefont {Wang}},\ }\href@noop {} {\bibfield
  {journal} {\bibinfo  {journal} {J. Am. Chem. Soc.}\ }\textbf {\bibinfo
  {volume} {113}},\ \bibinfo {pages} {2328} (\bibinfo {year}
  {1991})}\BibitemShut {NoStop}%
\bibitem [{\citenamefont {Mitzi}\ \emph {et~al.}(1995)\citenamefont {Mitzi},
  \citenamefont {Wang}, \citenamefont {Feild}, \citenamefont {Chess},\ and\
  \citenamefont {Guloy}}]{mitzi-1}%
  \BibitemOpen
  \bibfield  {author} {\bibinfo {author} {\bibfnamefont {D.~B.}\ \bibnamefont
  {Mitzi}}, \bibinfo {author} {\bibfnamefont {S.}~\bibnamefont {Wang}},
  \bibinfo {author} {\bibfnamefont {C.~A.}\ \bibnamefont {Feild}}, \bibinfo
  {author} {\bibfnamefont {C.~A.}\ \bibnamefont {Chess}}, \ and\ \bibinfo
  {author} {\bibfnamefont {A.~M.}\ \bibnamefont {Guloy}},\ }\href@noop {}
  {\bibfield  {journal} {\bibinfo  {journal} {Science}\ }\textbf {\bibinfo
  {volume} {267}},\ \bibinfo {pages} {1473} (\bibinfo {year}
  {1995})}\BibitemShut {NoStop}%
\bibitem [{\citenamefont {Liang}, \citenamefont {Mitzi},\ and\ \citenamefont
  {Prikas}(1998)}]{mitzi-2}%
  \BibitemOpen
  \bibfield  {author} {\bibinfo {author} {\bibfnamefont {K.}~\bibnamefont
  {Liang}}, \bibinfo {author} {\bibfnamefont {D.~B.}\ \bibnamefont {Mitzi}}, \
  and\ \bibinfo {author} {\bibfnamefont {M.~T.}\ \bibnamefont {Prikas}},\
  }\href@noop {} {\bibfield  {journal} {\bibinfo  {journal} {Chem. Mater.}\
  }\textbf {\bibinfo {volume} {10}},\ \bibinfo {pages} {403} (\bibinfo {year}
  {1998})}\BibitemShut {NoStop}%
\bibitem [{\citenamefont {Borriello}, \citenamefont {Cantele},\ and\
  \citenamefont {Ninno}(2008)}]{borriello-235214}%
  \BibitemOpen
  \bibfield  {author} {\bibinfo {author} {\bibfnamefont {I.}~\bibnamefont
  {Borriello}}, \bibinfo {author} {\bibfnamefont {G.}~\bibnamefont {Cantele}},
  \ and\ \bibinfo {author} {\bibfnamefont {D.}~\bibnamefont {Ninno}},\ }\href
  {\doibase 10.1103/PhysRevB.77.235214} {\bibfield  {journal} {\bibinfo
  {journal} {Phys. Rev. B}\ }\textbf {\bibinfo {volume} {77}},\ \bibinfo
  {pages} {235214} (\bibinfo {year} {2008})}\BibitemShut {NoStop}%
\bibitem [{\citenamefont {Mosconi}\ \emph {et~al.}(2013)\citenamefont
  {Mosconi}, \citenamefont {Amat}, \citenamefont {Nazeeruddin}, \citenamefont
  {Gr{\"a}tzel},\ and\ \citenamefont {De~Angelis}}]{mosconi-13902}%
  \BibitemOpen
  \bibfield  {author} {\bibinfo {author} {\bibfnamefont {E.}~\bibnamefont
  {Mosconi}}, \bibinfo {author} {\bibfnamefont {A.}~\bibnamefont {Amat}},
  \bibinfo {author} {\bibfnamefont {M.~K.}\ \bibnamefont {Nazeeruddin}},
  \bibinfo {author} {\bibfnamefont {M.}~\bibnamefont {Gr{\"a}tzel}}, \ and\
  \bibinfo {author} {\bibfnamefont {F.}~\bibnamefont {De~Angelis}},\ }\href
  {\doibase 10.1021/jp4048659} {\bibfield  {journal} {\bibinfo  {journal} {J.
  Phys. Chem. C}\ }\textbf {\bibinfo {volume} {117}},\ \bibinfo {pages} {13902}
  (\bibinfo {year} {2013})}\BibitemShut {NoStop}%
\bibitem [{\citenamefont {Quarti}\ \emph {et~al.}(2014)\citenamefont {Quarti},
  \citenamefont {Grancini}, \citenamefont {Mosconi}, \citenamefont {Bruno},
  \citenamefont {Ball}, \citenamefont {Lee}, \citenamefont {Snaith},
  \citenamefont {Petrozza},\ and\ \citenamefont {De~Angelis}}]{quarti-279}%
  \BibitemOpen
  \bibfield  {author} {\bibinfo {author} {\bibfnamefont {C.}~\bibnamefont
  {Quarti}}, \bibinfo {author} {\bibfnamefont {G.}~\bibnamefont {Grancini}},
  \bibinfo {author} {\bibfnamefont {E.}~\bibnamefont {Mosconi}}, \bibinfo
  {author} {\bibfnamefont {P.}~\bibnamefont {Bruno}}, \bibinfo {author}
  {\bibfnamefont {J.~M.}\ \bibnamefont {Ball}}, \bibinfo {author}
  {\bibfnamefont {M.~M.}\ \bibnamefont {Lee}}, \bibinfo {author} {\bibfnamefont
  {H.~J.}\ \bibnamefont {Snaith}}, \bibinfo {author} {\bibfnamefont
  {A.}~\bibnamefont {Petrozza}}, \ and\ \bibinfo {author} {\bibfnamefont
  {F.}~\bibnamefont {De~Angelis}},\ }\href {\doibase 10.1021/jz402589q}
  {\bibfield  {journal} {\bibinfo  {journal} {J. Phys. Chem. Lett.}\ }\textbf
  {\bibinfo {volume} {5}},\ \bibinfo {pages} {279} (\bibinfo {year}
  {2014})}\BibitemShut {NoStop}%
\bibitem [{\citenamefont {Brivio}, \citenamefont {Walker},\ and\ \citenamefont
  {Walsh}(2013)}]{brivio-042111}%
  \BibitemOpen
  \bibfield  {author} {\bibinfo {author} {\bibfnamefont {F.}~\bibnamefont
  {Brivio}}, \bibinfo {author} {\bibfnamefont {A.~B.}\ \bibnamefont {Walker}},
  \ and\ \bibinfo {author} {\bibfnamefont {A.}~\bibnamefont {Walsh}},\
  }\href@noop {} {\bibfield  {journal} {\bibinfo  {journal} {APL Mater.}\
  }\textbf {\bibinfo {volume} {1}},\ \bibinfo {pages} {042111} (\bibinfo {year}
  {2013})}\BibitemShut {NoStop}%
\bibitem [{\citenamefont {Yin}, \citenamefont {Shi},\ and\ \citenamefont
  {Yan}(2014)}]{yin-063903}%
  \BibitemOpen
  \bibfield  {author} {\bibinfo {author} {\bibfnamefont {W.-J.}\ \bibnamefont
  {Yin}}, \bibinfo {author} {\bibfnamefont {T.}~\bibnamefont {Shi}}, \ and\
  \bibinfo {author} {\bibfnamefont {Y.}~\bibnamefont {Yan}},\ }\href {\doibase
  http://dx.doi.org/10.1063/1.4864778} {\bibfield  {journal} {\bibinfo
  {journal} {Appl. Phys. Lett.}\ }\textbf {\bibinfo {volume} {104}},\ \bibinfo
  {eid} {063903} (\bibinfo {year} {2014})}\BibitemShut {NoStop}%
\bibitem [{\citenamefont {Filippetti}\ and\ \citenamefont
  {Mattoni}(2014)}]{filippetti-125203}%
  \BibitemOpen
  \bibfield  {author} {\bibinfo {author} {\bibfnamefont {A.}~\bibnamefont
  {Filippetti}}\ and\ \bibinfo {author} {\bibfnamefont {A.}~\bibnamefont
  {Mattoni}},\ }\href {http://link.aps.org/doi/10.1103/PhysRevB.89.125203}
  {\bibfield  {journal} {\bibinfo  {journal} {Phys. Rev. B}\ }\textbf {\bibinfo
  {volume} {89}},\ \bibinfo {pages} {125203} (\bibinfo {year}
  {2014})}\BibitemShut {NoStop}%
\bibitem [{\citenamefont {Even}\ \emph {et~al.}(2013)\citenamefont {Even},
  \citenamefont {Pedesseau}, \citenamefont {Jancu},\ and\ \citenamefont
  {Katan}}]{even-2999}%
  \BibitemOpen
  \bibfield  {author} {\bibinfo {author} {\bibfnamefont {J.}~\bibnamefont
  {Even}}, \bibinfo {author} {\bibfnamefont {L.}~\bibnamefont {Pedesseau}},
  \bibinfo {author} {\bibfnamefont {J.-M.}\ \bibnamefont {Jancu}}, \ and\
  \bibinfo {author} {\bibfnamefont {C.}~\bibnamefont {Katan}},\ }\href@noop {}
  {\bibfield  {journal} {\bibinfo  {journal} {J. Phys. Chem. Lett.}\ }\textbf
  {\bibinfo {volume} {4}},\ \bibinfo {pages} {2999} (\bibinfo {year}
  {2013})}\BibitemShut {NoStop}%
\bibitem [{\citenamefont {Giorgi}\ \emph {et~al.}(2013)\citenamefont {Giorgi},
  \citenamefont {Fujisawa}, \citenamefont {Segawa},\ and\ \citenamefont
  {Yamashita}}]{giorgi-4213}%
  \BibitemOpen
  \bibfield  {author} {\bibinfo {author} {\bibfnamefont {G.}~\bibnamefont
  {Giorgi}}, \bibinfo {author} {\bibfnamefont {J.-I.}\ \bibnamefont
  {Fujisawa}}, \bibinfo {author} {\bibfnamefont {H.}~\bibnamefont {Segawa}}, \
  and\ \bibinfo {author} {\bibfnamefont {K.}~\bibnamefont {Yamashita}},\
  }\href@noop {} {\bibfield  {journal} {\bibinfo  {journal} {J. Phys. Chem.
  Lett.}\ }\textbf {\bibinfo {volume} {4}},\ \bibinfo {pages} {4213} (\bibinfo
  {year} {2013})}\BibitemShut {NoStop}%
\bibitem [{\citenamefont {van Schilfgaarde}, \citenamefont {Kotani},\ and\
  \citenamefont {Faleev}(2006{\natexlab{a}})}]{mark06qsgw}%
  \BibitemOpen
  \bibfield  {author} {\bibinfo {author} {\bibfnamefont {M.}~\bibnamefont {van
  Schilfgaarde}}, \bibinfo {author} {\bibfnamefont {T.}~\bibnamefont {Kotani}},
  \ and\ \bibinfo {author} {\bibfnamefont {S.}~\bibnamefont {Faleev}},\
  }\href@noop {} {\bibfield  {journal} {\bibinfo  {journal} {Phys. Rev. Lett.}\
  }\textbf {\bibinfo {volume} {96}},\ \bibinfo {eid} {226402} (\bibinfo {year}
  {2006}{\natexlab{a}})}\BibitemShut {NoStop}%
\bibitem [{\citenamefont {Vidal}\ \emph {et~al.}(2010)\citenamefont {Vidal},
  \citenamefont {Botti}, \citenamefont {Olsson}, \citenamefont {Guillemoles},\
  and\ \citenamefont {Reining}}]{vidal-056401}%
  \BibitemOpen
  \bibfield  {author} {\bibinfo {author} {\bibfnamefont {J.}~\bibnamefont
  {Vidal}}, \bibinfo {author} {\bibfnamefont {S.}~\bibnamefont {Botti}},
  \bibinfo {author} {\bibfnamefont {P.}~\bibnamefont {Olsson}}, \bibinfo
  {author} {\bibfnamefont {J.-F. m.~c.}\ \bibnamefont {Guillemoles}}, \ and\
  \bibinfo {author} {\bibfnamefont {L.}~\bibnamefont {Reining}},\ }\href
  {\doibase 10.1103/PhysRevLett.104.056401} {\bibfield  {journal} {\bibinfo
  {journal} {Phys. Rev. Lett.}\ }\textbf {\bibinfo {volume} {104}},\ \bibinfo
  {pages} {056401} (\bibinfo {year} {2010})}\BibitemShut {NoStop}%
\bibitem [{\citenamefont {Perdew}\ \emph {et~al.}(2008)\citenamefont {Perdew},
  \citenamefont {Ruzsinszky}, \citenamefont {Csonka}, \citenamefont {Vydrov},
  \citenamefont {Scuseria}, \citenamefont {Constantin}, \citenamefont {Zhou},\
  and\ \citenamefont {Burke}}]{pbesol}%
  \BibitemOpen
  \bibfield  {author} {\bibinfo {author} {\bibfnamefont {J.~P.}\ \bibnamefont
  {Perdew}}, \bibinfo {author} {\bibfnamefont {A.}~\bibnamefont {Ruzsinszky}},
  \bibinfo {author} {\bibfnamefont {G.~I.}\ \bibnamefont {Csonka}}, \bibinfo
  {author} {\bibfnamefont {O.~A.}\ \bibnamefont {Vydrov}}, \bibinfo {author}
  {\bibfnamefont {G.~E.}\ \bibnamefont {Scuseria}}, \bibinfo {author}
  {\bibfnamefont {L.~A.}\ \bibnamefont {Constantin}}, \bibinfo {author}
  {\bibfnamefont {X.}~\bibnamefont {Zhou}}, \ and\ \bibinfo {author}
  {\bibfnamefont {K.}~\bibnamefont {Burke}},\ }\href@noop {} {\bibfield
  {journal} {\bibinfo  {journal} {Phys. Rev. Lett.}\ }\textbf {\bibinfo
  {volume} {100}},\ \bibinfo {pages} {136406} (\bibinfo {year}
  {2008})}\BibitemShut {NoStop}%
\bibitem [{\citenamefont {Usuda}\ \emph {et~al.}(2004)\citenamefont {Usuda},
  \citenamefont {Hamada}, \citenamefont {Siraishi},\ and\ \citenamefont
  {Oshiyama}}]{Usuda04}%
  \BibitemOpen
  \bibfield  {author} {\bibinfo {author} {\bibfnamefont {M.}~\bibnamefont
  {Usuda}}, \bibinfo {author} {\bibfnamefont {H.}~\bibnamefont {Hamada}},
  \bibinfo {author} {\bibfnamefont {K.}~\bibnamefont {Siraishi}}, \ and\
  \bibinfo {author} {\bibfnamefont {A.}~\bibnamefont {Oshiyama}},\ }\href@noop
  {} {\bibfield  {journal} {\bibinfo  {journal} {Jpn. J. Appl. Phys. Lett.
  (part 2)}\ }\textbf {\bibinfo {volume} {43}},\ \bibinfo {pages} {L407}
  (\bibinfo {year} {2004})}\BibitemShut {NoStop}%
\bibitem [{\citenamefont {Yamada}\ \emph {et~al.}(2014)\citenamefont {Yamada},
  \citenamefont {Nakamura}, \citenamefont {Endo}, \citenamefont {Wakamiya},\
  and\ \citenamefont {Kanemitsu}}]{yamada-032302}%
  \BibitemOpen
  \bibfield  {author} {\bibinfo {author} {\bibfnamefont {Y.}~\bibnamefont
  {Yamada}}, \bibinfo {author} {\bibfnamefont {T.}~\bibnamefont {Nakamura}},
  \bibinfo {author} {\bibfnamefont {M.}~\bibnamefont {Endo}}, \bibinfo {author}
  {\bibfnamefont {A.}~\bibnamefont {Wakamiya}}, \ and\ \bibinfo {author}
  {\bibfnamefont {Y.}~\bibnamefont {Kanemitsu}},\ }\href
  {http://stacks.iop.org/1882-0786/7/i=3/a=032302} {\bibfield  {journal}
  {\bibinfo  {journal} {Appl. Phys. Expr.}\ }\textbf {\bibinfo {volume} {7}},\
  \bibinfo {pages} {032302} (\bibinfo {year} {2014})}\BibitemShut {NoStop}%
\bibitem [{con()}]{convergence}%
  \BibitemOpen
  \href@noop {} {}\bibinfo {note} {Careful convergence checks were made of the
  eigenfunction and product basis. A 4$\times$4$\times$4 $k$-mesh was used for
  the M-PbI$_3$ compounds. A 1-shot calculation with a 6$\times$6$\times$6 mesh
  as perturbation to the QS\emph{GW} result showed that the $k$-converged gap
  is about 0.1\,eV larger than what is reported in Table I.}\BibitemShut
  {Stop}%
\bibitem [{\citenamefont {Poglitsch}\ and\ \citenamefont
  {Weber}(1987)}]{poglitsch1987dynamic}%
  \BibitemOpen
  \bibfield  {author} {\bibinfo {author} {\bibfnamefont {A.}~\bibnamefont
  {Poglitsch}}\ and\ \bibinfo {author} {\bibfnamefont {D.}~\bibnamefont
  {Weber}},\ }\href@noop {} {\bibfield  {journal} {\bibinfo  {journal} {J.
  Chem. Phys.}\ }\textbf {\bibinfo {volume} {87}},\ \bibinfo {pages} {6373}
  (\bibinfo {year} {1987})}\BibitemShut {NoStop}%
\bibitem [{\citenamefont {van Schilfgaarde}, \citenamefont {Kotani},\ and\
  \citenamefont {Faleev}(2006{\natexlab{b}})}]{mark06adeq}%
  \BibitemOpen
  \bibfield  {author} {\bibinfo {author} {\bibfnamefont {M.}~\bibnamefont {van
  Schilfgaarde}}, \bibinfo {author} {\bibfnamefont {T.}~\bibnamefont {Kotani}},
  \ and\ \bibinfo {author} {\bibfnamefont {S.~V.}\ \bibnamefont {Faleev}},\
  }\href@noop {} {\bibfield  {journal} {\bibinfo  {journal} {Phys. Rev. B}\
  }\textbf {\bibinfo {volume} {74}},\ \bibinfo {pages} {245125} (\bibinfo
  {year} {2006}{\natexlab{b}})}\BibitemShut {NoStop}%
\bibitem [{\citenamefont {De~Wolf}\ \emph {et~al.}(2014)\citenamefont
  {De~Wolf}, \citenamefont {Holovsky}, \citenamefont {Moon}, \citenamefont
  {L{\"o}per}, \citenamefont {Niesen}, \citenamefont {Ledinsky}, \citenamefont
  {Haug}, \citenamefont {Yum},\ and\ \citenamefont {Ballif}}]{wolf-2014}%
  \BibitemOpen
  \bibfield  {author} {\bibinfo {author} {\bibfnamefont {S.}~\bibnamefont
  {De~Wolf}}, \bibinfo {author} {\bibfnamefont {J.}~\bibnamefont {Holovsky}},
  \bibinfo {author} {\bibfnamefont {S.-J.}\ \bibnamefont {Moon}}, \bibinfo
  {author} {\bibfnamefont {P.}~\bibnamefont {L{\"o}per}}, \bibinfo {author}
  {\bibfnamefont {B.}~\bibnamefont {Niesen}}, \bibinfo {author} {\bibfnamefont
  {M.}~\bibnamefont {Ledinsky}}, \bibinfo {author} {\bibfnamefont {F.-J.}\
  \bibnamefont {Haug}}, \bibinfo {author} {\bibfnamefont {J.-H.}\ \bibnamefont
  {Yum}}, \ and\ \bibinfo {author} {\bibfnamefont {C.}~\bibnamefont {Ballif}},\
  }\href {\doibase 10.1021/jz500279b} {\bibfield  {journal} {\bibinfo
  {journal} {J. Phys. Chem. Lett.}\ }\textbf {\bibinfo {volume} {5}},\ \bibinfo
  {pages} {1035} (\bibinfo {year} {2014})}\BibitemShut {NoStop}%
\bibitem [{\citenamefont {Onoda-Yamamuro}, \citenamefont {Matsuo},\ and\
  \citenamefont {Suga}(1992)}]{onoda-935}%
  \BibitemOpen
  \bibfield  {author} {\bibinfo {author} {\bibfnamefont {N.}~\bibnamefont
  {Onoda-Yamamuro}}, \bibinfo {author} {\bibfnamefont {T.}~\bibnamefont
  {Matsuo}}, \ and\ \bibinfo {author} {\bibfnamefont {H.}~\bibnamefont
  {Suga}},\ }\href@noop {} {\bibfield  {journal} {\bibinfo  {journal} {J. Phys.
  Chem. Sol.}\ }\textbf {\bibinfo {volume} {53}},\ \bibinfo {pages} {935}
  (\bibinfo {year} {1992})}\BibitemShut {NoStop}%
\bibitem [{\citenamefont {Frost}\ \emph {et~al.}(2014)\citenamefont {Frost},
  \citenamefont {Butler}, \citenamefont {Brivio}, \citenamefont {Hendon},
  \citenamefont {van Schilfgaarde},\ and\ \citenamefont {Walsh}}]{frost-2014}%
  \BibitemOpen
  \bibfield  {author} {\bibinfo {author} {\bibfnamefont {J.~M.}\ \bibnamefont
  {Frost}}, \bibinfo {author} {\bibfnamefont {K.~T.}\ \bibnamefont {Butler}},
  \bibinfo {author} {\bibfnamefont {F.}~\bibnamefont {Brivio}}, \bibinfo
  {author} {\bibfnamefont {C.~H.}\ \bibnamefont {Hendon}}, \bibinfo {author}
  {\bibfnamefont {M.}~\bibnamefont {van Schilfgaarde}}, \ and\ \bibinfo
  {author} {\bibfnamefont {A.}~\bibnamefont {Walsh}},\ }\href {\doibase
  10.1021/nl500390f} {\bibfield  {journal} {\bibinfo  {journal} {Nano Lett.}\
  }\textbf {\bibinfo {volume} {In Press}} (\bibinfo {year} {2014}),\
  10.1021/nl500390f}\BibitemShut {NoStop}%
\bibitem [{\citenamefont {Watson}, \citenamefont {Oliver},\ and\ \citenamefont
  {Parker}(1997)}]{metadise}%
  \BibitemOpen
  \bibfield  {author} {\bibinfo {author} {\bibfnamefont {G.~W.}\ \bibnamefont
  {Watson}}, \bibinfo {author} {\bibfnamefont {P.~M.}\ \bibnamefont {Oliver}},
  \ and\ \bibinfo {author} {\bibfnamefont {S.~C.}\ \bibnamefont {Parker}},\
  }\href@noop {} {\bibfield  {journal} {\bibinfo  {journal} {Phys. Chem.
  Mater.}\ }\textbf {\bibinfo {volume} {25}},\ \bibinfo {pages} {70} (\bibinfo
  {year} {1997})}\BibitemShut {NoStop}%
\bibitem [{\citenamefont {Butler}\ \emph {et~al.}(2014)\citenamefont {Butler},
  \citenamefont {Buckeridge}, \citenamefont {Catlow},\ and\ \citenamefont
  {Walsh}}]{butler-115320}%
  \BibitemOpen
  \bibfield  {author} {\bibinfo {author} {\bibfnamefont {K.~T.}\ \bibnamefont
  {Butler}}, \bibinfo {author} {\bibfnamefont {J.}~\bibnamefont {Buckeridge}},
  \bibinfo {author} {\bibfnamefont {C.~R.~A.}\ \bibnamefont {Catlow}}, \ and\
  \bibinfo {author} {\bibfnamefont {A.}~\bibnamefont {Walsh}},\ }\href@noop {}
  {\bibfield  {journal} {\bibinfo  {journal} {Phys. Rev. B.}\ }\textbf
  {\bibinfo {volume} {89}},\ \bibinfo {pages} {115320} (\bibinfo {year}
  {2014})}\BibitemShut {NoStop}%
\bibitem [{\citenamefont {Walsh}\ and\ \citenamefont
  {Butler}(2014)}]{butler-2014}%
  \BibitemOpen
  \bibfield  {author} {\bibinfo {author} {\bibfnamefont {A.}~\bibnamefont
  {Walsh}}\ and\ \bibinfo {author} {\bibfnamefont {K.~T.}\ \bibnamefont
  {Butler}},\ }\href@noop {} {\bibfield  {journal} {\bibinfo  {journal} {Acc.
  Chem. Res.}\ }\textbf {\bibinfo {volume} {47}},\ \bibinfo {pages} {364}
  (\bibinfo {year} {2014})}\BibitemShut {NoStop}%
\bibitem [{\citenamefont {Park}(2013)}]{park-2423}%
  \BibitemOpen
  \bibfield  {author} {\bibinfo {author} {\bibfnamefont {N.-G.}\ \bibnamefont
  {Park}},\ }\href@noop {} {\bibfield  {journal} {\bibinfo  {journal} {J. Phys.
  Chem. Lett.}\ }\textbf {\bibinfo {volume} {4}},\ \bibinfo {pages} {2423}
  (\bibinfo {year} {2013})}\BibitemShut {NoStop}%
\bibitem [{\citenamefont {Aryasetiawan}\ and\ \citenamefont
  {Biermann}(2008)}]{Aryasetiawan08}%
  \BibitemOpen
  \bibfield  {author} {\bibinfo {author} {\bibfnamefont {F.}~\bibnamefont
  {Aryasetiawan}}\ and\ \bibinfo {author} {\bibfnamefont {S.}~\bibnamefont
  {Biermann}},\ }\href@noop {} {\bibfield  {journal} {\bibinfo  {journal}
  {Phys. Rev. Lett.}\ }\textbf {\bibinfo {volume} {100}},\ \bibinfo {pages}
  {116402} (\bibinfo {year} {2008})}\BibitemShut {NoStop}%
\end{thebibliography}%

\end{document}